\documentclass[aps,prl,reprint,showpacs,amsmath,amssymb,superscriptaddress,floatfix]{revtex4-1}
\usepackage{amsmath}
\usepackage{amssymb}
\usepackage{bm}
\usepackage{graphics}
\usepackage{float}
\usepackage{color}
\usepackage{bbold}
\usepackage{graphicx}
\usepackage{ifpdf}
\usepackage{epstopdf}
\usepackage{bbold}
\usepackage{caption}
\usepackage{subcaption}
\begin{document}
\title{Metal-insulator transition in a sliding Luttinger liquid with line defects}



\affiliation{I. Institut f\"ur Theoretische Physik, Universit\"at Hamburg, Hamburg, Germany}
\affiliation{Shamoon College of Engineering, Bialik/Basel St., Beer-Sheva 84100, Israel}
\affiliation{Nonlinearity and Complexity Research Group, School of Engineering \& Applied Science, Aston University, Birmingham B4 7ET, United Kingdom}
\affiliation{Institute of Fundamental and Frontier Sciences, University of Electronic Science and Technology of China, Chengdu 610054, People's Republic of China}
\affiliation{Center for Theoretical Physics of Complex Systems, Institute for Basic Science, Daejeon, South Korea}

\author{A. L. Chudnovskiy}
\affiliation{I. Institut f\"ur Theoretische Physik, Universit\"at Hamburg, Hamburg, Germany}
\affiliation{Center for Theoretical Physics of Complex Systems, Institute for Basic Science, Daejeon, South Korea}

\author{V. Kagalovsky}
\affiliation{Shamoon College of Engineering, Bialik/Basel St., Beer-Sheva 84100, Israel}
\affiliation{Institute of Fundamental and Frontier Sciences, University of Electronic Science and Technology of China, Chengdu 610054, People's Republic of China}
\affiliation{Center for Theoretical Physics of Complex Systems, Institute for Basic Science, Daejeon, South Korea}

\author{I. V. Yurkevich}
\affiliation{Nonlinearity and Complexity Research Group, School of Engineering \& Applied Science, Aston University, Birmingham B4 7ET, United Kingdom}
\affiliation{Institute of Fundamental and Frontier Sciences, University of Electronic Science and Technology of China, Chengdu 610054, People's Republic of China}
\affiliation{Center for Theoretical Physics of Complex Systems, Institute for Basic Science, Daejeon, South Korea}

\begin{abstract}
We investigate the effect of both strong and weak potential scattering caused by local impurities
and extended (line) defects in the array of Luttinger liquid wires. We find that in both cases a finite
range inter-wire interaction stabilises metallic state. Based on calculation of the scaling dimensions
of one-particle scattering operators, we construct the phase diagram for low-temperature transport
along the array. We find that unlike local impurity case where only conducting and insulating states 
are realised (metal-insulator transition driven by interactions), the extended line defects may
bring the system into a mixed state where conducting or insulating behavior can be observed
depending on bare strength of the scatterer (metal-insulator transition driven by disorder).
\end{abstract}

\date{\today}


\maketitle

\section{Introduction}
Since the prediction of the Anderson metal-insulator transition \cite{AndersonMIT,AALR}, the effects of interactions on this quantum phase transition  remain a puzzling problem, despite a huge body of theoretical and experimental investigations. For the weak localization, in the diffusive regime, inter-particle interactions are known to favour further localization of the system \cite{Alt_Ar_Lee80}. At the same time, experiments on very clean two-dimensional systems show evidence of a metal-insulator transition driven by change of interaction strength  (for a review, see Ref. \cite{Kravchenko1}). Another important theoretical break-through in the understanding of the role of interaction is the discovery of the many-body delocalization, which occurs due to interactions in the isolated system at temperatures above the critical one \cite{BAA}. The theoretical description of interaction for the generic disordered electron system requires to go beyond the perturbation theory, thus posing a great theoretical challenge. At the same time, the interactions can be taken into account nonperturbatively and to great extent exactly in one-dimensional electron systems in frame of the Luttinger liquid (LL) theory.  It is therefore tempting to attack the problem of the effect of interactions for Anderson localization by making use of the Luttinger liquid technique.

Recent advances in study of strongly correlated systems have led to wider search for exotic non-Fermi-liquid states in condensed matter systems. In particular, the quest for edge states protected against disorder has started. These states are protected by a symmetry that forbids perturbations potentially dangerous for the phase stability. Another option is the renormalisation of dangerous perturbations such that they become suppressed at low temperatures and vanish in zero-temperature regime leading to a metallic   zero-temperature state protected by interactions. One of the promising models providing rich non-Fermi-liquid physics is an anisotropic system consisting of array of coupled one-dimensional wires \cite{TK,BN,Po}. This model was used for construction of integer \cite{Sondhi} and fractional quantum Hall states \cite{Kane2002}. Sliding phases in classical $XY$ models \cite{XY}, smectic metals \cite{smectic} and many other exotic states are all described by the sliding Luttinger liquid (sLL) model \cite{sLL,SLL,Santos15,Scheidl02}. The sLL is a model of an array of parallel LL wires with charge-density interactions between the different LL wires but without the single-particle hopping between them. Interwire interactions may provide the charge density fluctuations that destroy the pinning of the charge density wave by the disorder scattering, thus suppressing the Anderson localization in the array \cite{smectic,sLL,SLL,Scheidl02}.

Infinite arrays of wires and few-channel liquids have been investigated for decades. It was established that to stay quasi-one-dimensional at zero temperature requires very special form of inter-wire interactions \cite{SLL}. For a smooth generic interaction the system falls into either pinned charge density wave or two-dimensional superconducting state. The stability of sLL phase against disorder is another wide area of research \cite{TK,Kane2002,XY,smectic,sLL,Scheidl02,SLL,Santos15}. In particular, the effect of scattering by local impurities in sLL was investigated in Ref. \cite{Scheidl02}. As it is shown below, our results on scattering by local impurities agree with those of Ref. \cite{Scheidl02}.  A single impurity embedded into a LL  and a continuous disorder in LL \cite{GS} are known to crucially affect single channel LL transport. The renormalisation group (RG) analysis shows that strictly 1D transport of repulsive electrons is completely blocked by a single impurity at zero temperature \cite{KF}. The generalisation of the RG-analysis to multichannel problem was initiated by the series of papers \cite{TK,Sondhi,Kane2002,XY,smectic,Scheidl02,sLL,SLL} and further advanced in Refs.  \cite{IVY,KLY,IVYnew,Santos15}.

\begin{figure}[H]
\centering
\includegraphics[width=0.7 \linewidth]{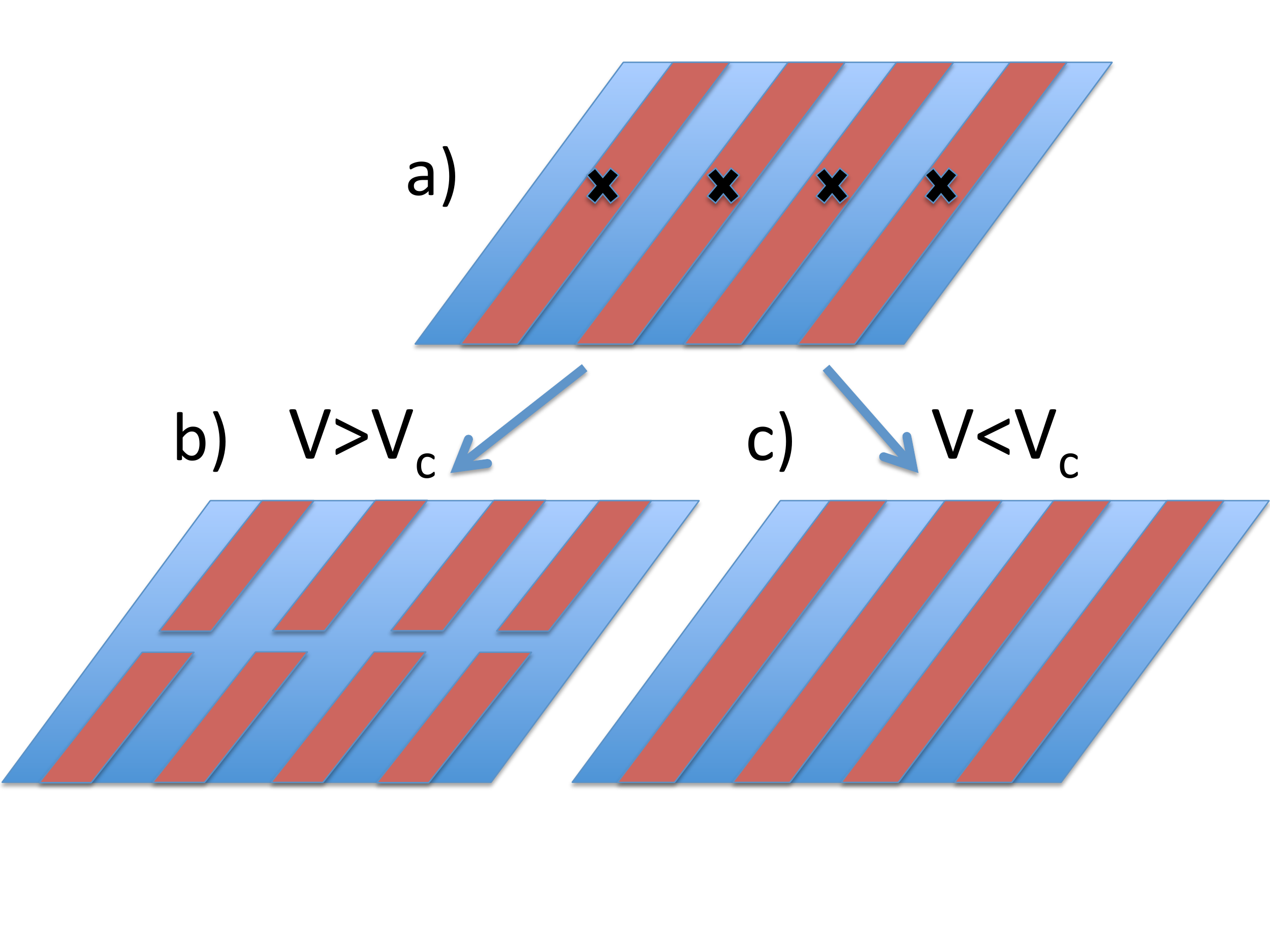}
\includegraphics[width=0.7 \linewidth]{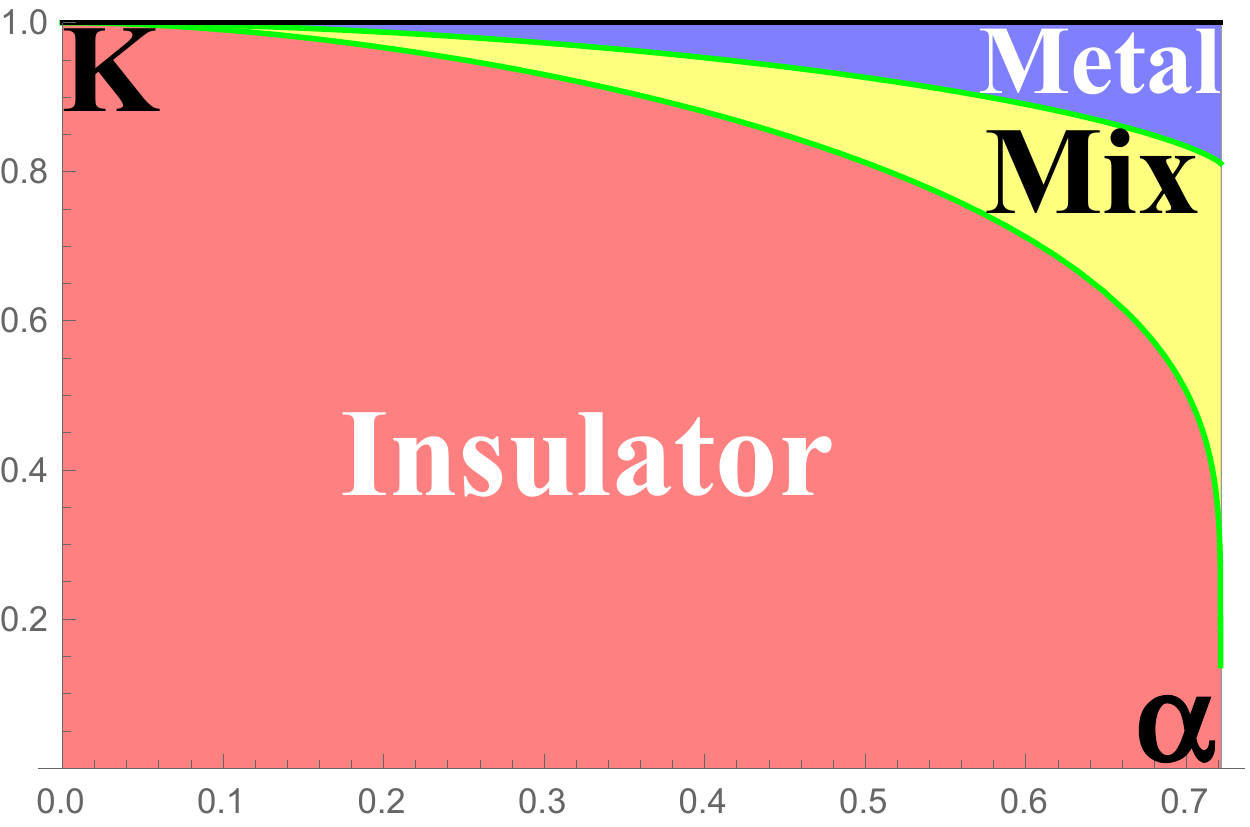}
\caption{Upper panel: a) Array of quantum wires with a line defect. b) Effective low-energy picture  for the back-scattering relevant, (insulating state, $V>V_c$ in the mixed state); c) Effective low-energy picture for the back-scattering irrelevant (metallic state, $V<V_c$ in the mixed state).
Lower panel: Phase diagram for a linear defect model for an infinite radius of inter-wire interaction (inter-wire radius of interaction parameter $\beta=1$). The state is specified by the inter-wire interaction strength $\alpha$ (horizontal axis), and the intra-wire Luttinger parameter $K$ (vertical axis). The existence of a mixed state (yellow, labeled “mix”) signals dependence on the initial (bare) strength of  scattering, leading to a disorder driven metal (blue) - insulator (red) transition.}
\label{figMixedState}
\end{figure}

In this paper, we perform the analysis of one-particle back-scattering by the potential impurities in the two-dimensional (2D) array of parallel quantum wires. Nonperturbative treatment of interactions allows exact evaluation of scaling dimensions of the scattering operators facilitating conclusions on the relevance/irrelevance of a given scattering channel. Based on the analysis of scaling dimensions, we show, that a
line of defects going through the whole array brings the system into a \emph{mixed} state, where the back-scattering may be relevant of irrelevant depending on its strength. The low temperature phase diagram of an array with a line defect is shown in Fig. \ref{figMixedState}. In the case of point defects, the phase diagram  exhibits a localized state, where the back-scattering is relevant, and a metallic state with back-scattering is irrelevant.

Since our phase diagrams are based on the analysis of scaling dimensions of single impurity scattering operators, the results obtained describe an array of wires with low density of multiple impurities at moderate temperatures when thermal length is smaller than the mean free path, $l=v_{\rm F}\tau$, and, therefore, each impurity renormalises individually in this high-temperature regime, $T\tau\gg 1$. We also assume that the temperatures $T_{*}$, when various inter-channel hybridisations, such as  single-particle hopping, charge density wave (CDW) and Josephson couplings, may become relevant, are low enough and are not reached. This temperatures are known to be defined by the bare couplings, $J_{*}$, the bandwidth $D$, and the scaling dimension, $\Delta_{*}$, of the most dangerous hybridization term: $T_{*}\sim D\,(J_{*}/D)^{\alpha}$ with the exponent $\alpha=(2-\Delta_{*})^{-1}$. In materials where overlap between electron wave functions that belong to the adjacent wires is small, the bare values of single-particle and correlated-pair inter-wire (Josephson) hopping are also small. The CDW-coupling is proportional to the Fourier transform of the Coulomb potential at $2k_{\rm F}$ and its bare value is small for a smooth (on the scale of the Fermi wave length) potentials. Under the assumption that coupling $J_{*}$ is much smaller that the bare impurity strength, there is a wide temperature range $T\gg T_{*},\tau^{-1}$, where we may safely use model of interacting but not hybridised wires (channels) with disorder modelled by individual impurities.

\section{The model}
\begin{figure}[H]
\centering
\includegraphics[width=0.8 \linewidth]{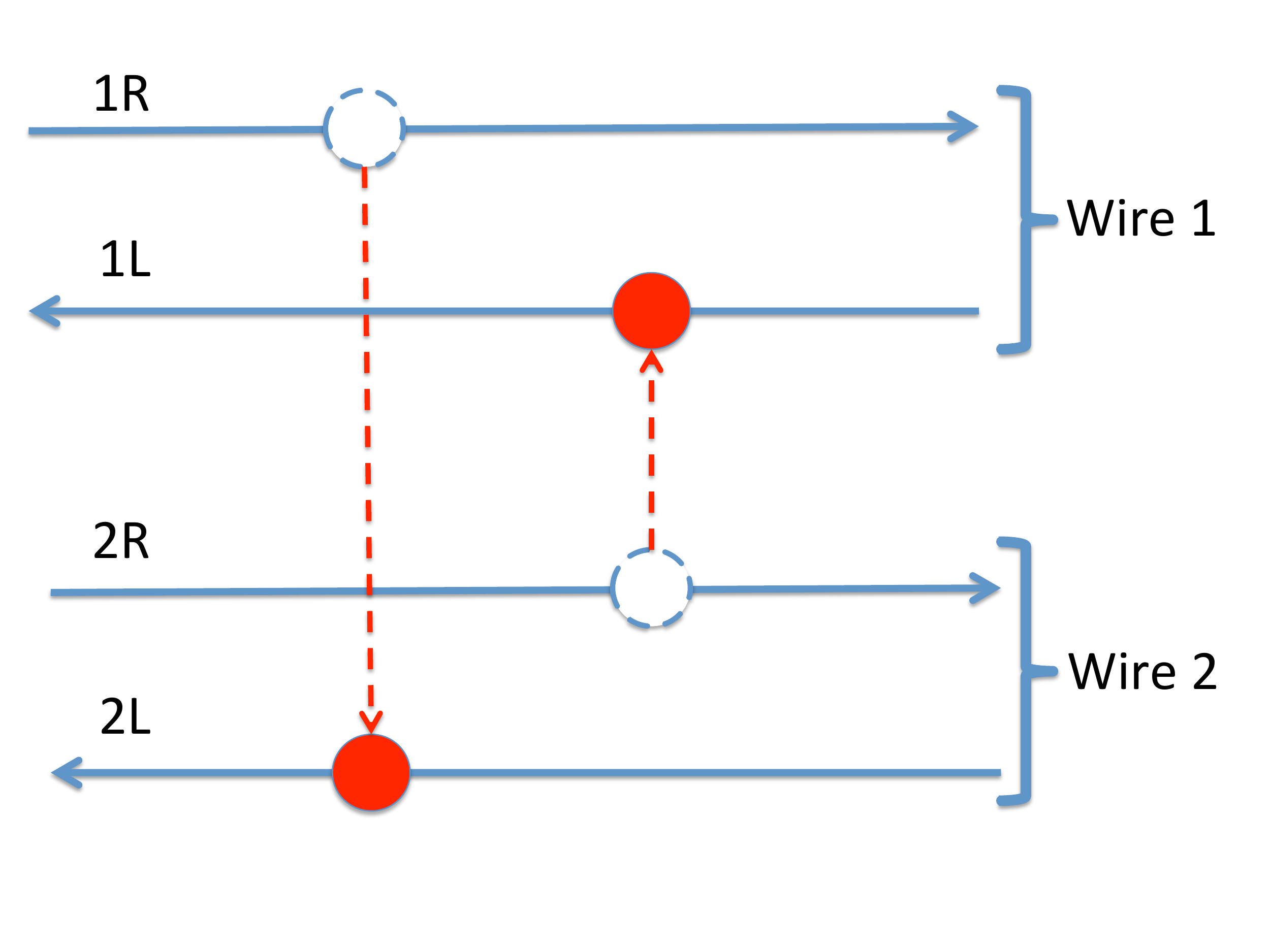}
\caption{Interwire backscattering operator described by the vectors ${\bf n}_{\mathrm{in}}=(1, 1)^T$, ${\bf n}_{\mathrm{out}}=(1, 1)^T$, ${\bf j}_{\mathrm{in}}=(1, 1)^T$, ${\bf j}_{\mathrm{out}}=(-1, -1)^T$.}
\label{figScattOP}
\end{figure}

We consider an array of identical parallel  LL wires. The bosonised action of the system with density-density interactions reads
\begin{equation}\label{H0}
H_0=\frac{v}{8\pi}\,\left[\partial_x{\bm\theta}^{\rm T}\,\left(1+{\hat g}\right)\,\partial_x{\bm\theta}+
\partial_x{\bm\phi}^{\rm T}\,\partial_x{\bm\phi}\right]\,,
\end{equation}
where fields ${\bm\theta}^{\rm T}=(\theta_1\,,...\,,\theta_N),$ and ${\bm\phi}^{\rm T}=(\phi_1\,,...\,,\phi_N)$
parameterise the density, $\rho_i=\partial_x\theta_i/2\pi$, and the current, $j_i=\partial_x\phi_i/2\pi$, in the $i$-th wire.
The density-density interactions are represented by the matrix ${\hat g}$ with elements $g_{ij}=g_{|i-j|}$ describing interaction between wires arranged into an array of 1D LLs. The technique for calculation of the scaling dimension of an operator describing multi-particle backscattering
by a single impurity was developed in the series of papers \cite{TK,Sondhi,Kane2002,XY,smectic,Scheidl02,sLL,SLL} and recently formalised in \cite{IVY,KLY,IVYnew,Santos15} and we refer the readers interested in the technical details to the latter. The following expression for the scaling dimension can be derived:
\begin{eqnarray}\label{Delta}
\Delta &=&\frac{1}{2}({\bf j}_{\rm in}-{\bf j}_{\rm out})^{\rm T}\,{\hat K}\,({\bf j}_{\rm in}-{\bf j}_{\rm out})\\\nonumber
&+&\frac{1}{2}({\bf n}_{\rm in}-{\bf n}_{\rm out})^{\rm T}\,{\hat K}^{-1}\,({\bf n}_{\rm in}-{\bf n}_{\rm out})\,.
\end{eqnarray}
Here the integer-valued vectors of in- and out-going particles numbers, ${\bf n}_{\mu}$, and currents, ${\bf j}_{\mu}$,
$$
{\bf n}_{\mu}={\bf n}^{\rm R}_{\mu}+{\bf n}^{\rm L}_{\mu}\,,\quad {\bf j}_{\mu}={\bf n}^{\rm R}_{\mu}-{\bf n}^{\rm L}_{\mu}\,,
\quad \mu={\rm in}\,,{\rm out}\,,
$$
describe multiplicity of the scattering: $n^{\eta}_{{\rm in}\,i}$ are the numbers of right- or left moving ($\eta=R,L$) particles annihilated  in the $i$-th channel and $n^{\eta}_{{\rm out}\,j}$ are the numbers of particles created in the $j$-th channel (see Fig. \ref{figScattOP} for an example). These integer numbers are constrained by current and particle number conservations:
\begin{eqnarray}
\begin{array}{c}
N_{\rm in}=N_{\rm out}\,,\\
J_{\rm in}=J_{\rm out}\,,
\end{array}\quad
\begin{array}{c}
N_{\mu}=\sum_{i}n_{{\mu}\,i}\,,\\
 J_{\mu}=\sum_{i}j_{{\mu}\,i}\,.
\end{array}
\end{eqnarray}
In the absence of inter-particle current-current interaction, the case considered in this paper, the Luttinger ${\hat K}$-matrix in Eq.~\ref{Delta} is given by (for details see Refs. \cite{KLY,IVYnew}):
\begin{equation}
{\hat K}=\left[1+{\hat g}\right]^{-1/2}\,.
\end{equation}
Since the matrix ${\hat g}$ is defined on a one-dimensional lattice and translation invariant, the explicit expression for matrix elements, $K_{ij}=K_n$ with $n=|i-j|$, of the Luttinger ${\hat K}$-matrix can be written as
\begin{equation}
K_n=\int_{0}^{\pi}\frac{{\rm d}q}{\pi}\,\frac{\cos qn}{\sqrt{1+g_q}}\,,\quad
g_q=\sum_{n=-\infty}^{\infty}\,g_n\,\cos qn\,.
\end{equation}
We assume that the Coulomb interaction in 2D array is screened by surrounding metallic gates, which leads to exponential decay of interaction with distance. Therefore the interaction  can be described by the following expression
\begin{eqnarray}\label{scrCoulomb}
g_{n=0}=g_0\,,\quad g_{n\neq 0}=g'\,|n|^{-1}\,e^{-\kappa |n|}\,,
\end{eqnarray}
where we introduced the transverse (inter-wire) screening parameter $\beta=\exp{(-\kappa)}$ with $\kappa^{-1}$ being the screening length measured in units of the inter-wire distance. Then the discrete Fourier transform of the transverse interaction is given by
\begin{equation}
g_q=g_0-g'\,\ln\left[1+\beta^2-2\beta\,\cos q\right].
\end{equation}
Finally, restoring standard Luttinger parameter $K=(1+g_0)^{-1/2}$ describing single wire without surrounding (i.e. $g'=0$), the Luttinger matrix can be parameterised as
\begin{eqnarray}\label{K}
K_n&=&K\,\int_{0}^{\pi}\frac{{\rm d}q}{\pi}\,\frac{\cos qn}{r_q}\,,\\
r_q&=&\sqrt{1-\alpha\,\ln\left[1+\beta^2-2\beta\,\cos q\right]}\,,\nonumber
\end{eqnarray}
where the parameter $\alpha=g'/(1+g_0)>0$ describes the relative strength of inter-channel repulsion.


\begin{figure*}
\centering
\includegraphics[width=.4\linewidth]{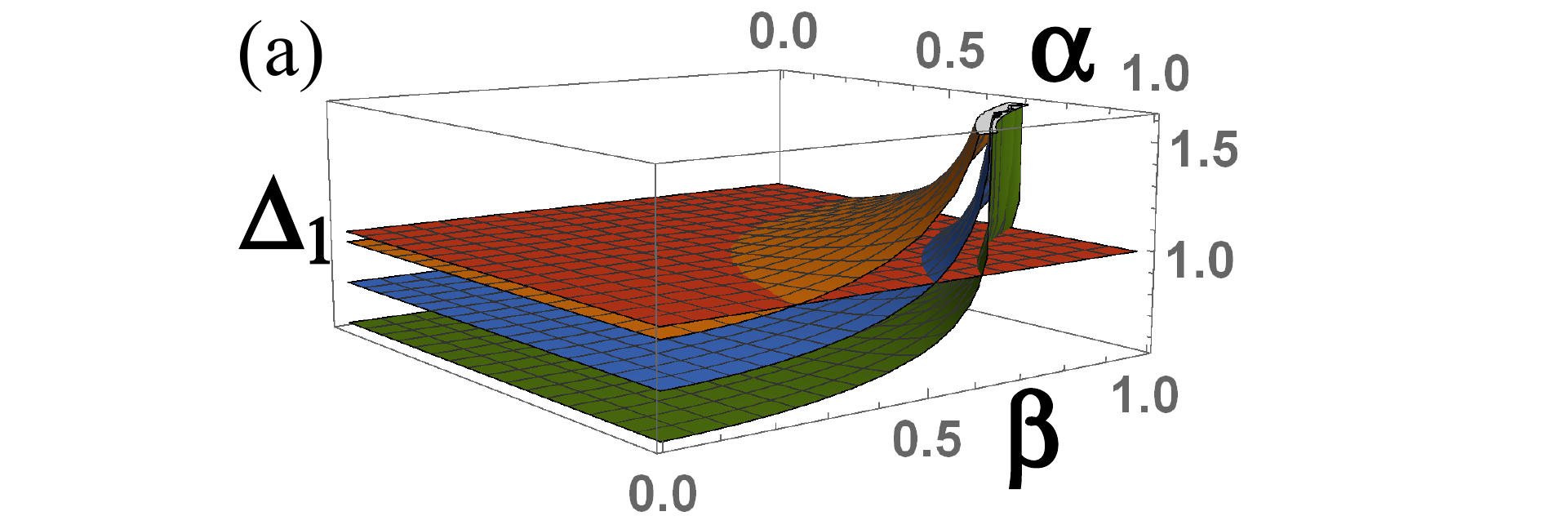}\qquad
\includegraphics[width=.15\linewidth]{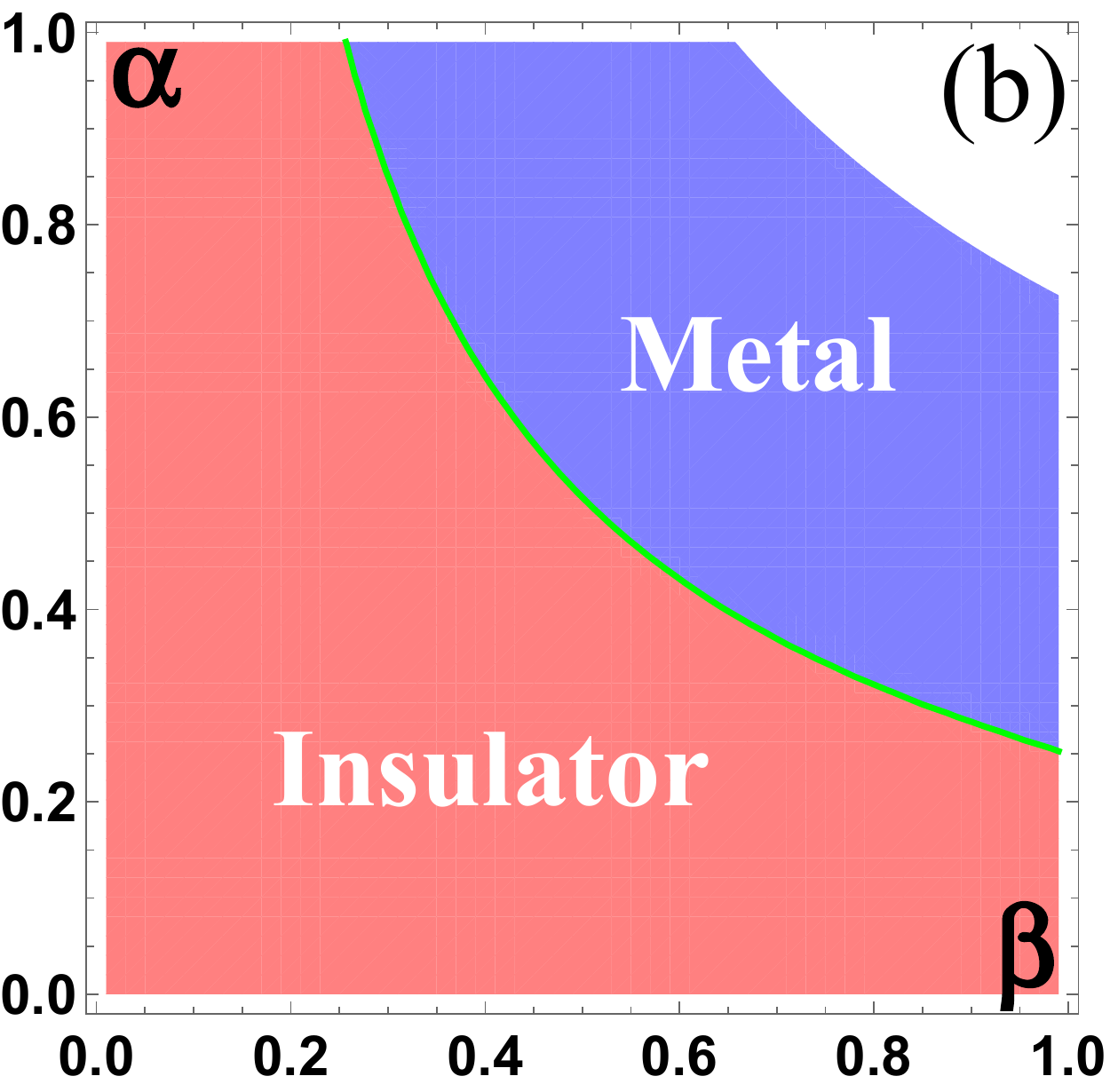}\quad
\includegraphics[width=.15\linewidth]{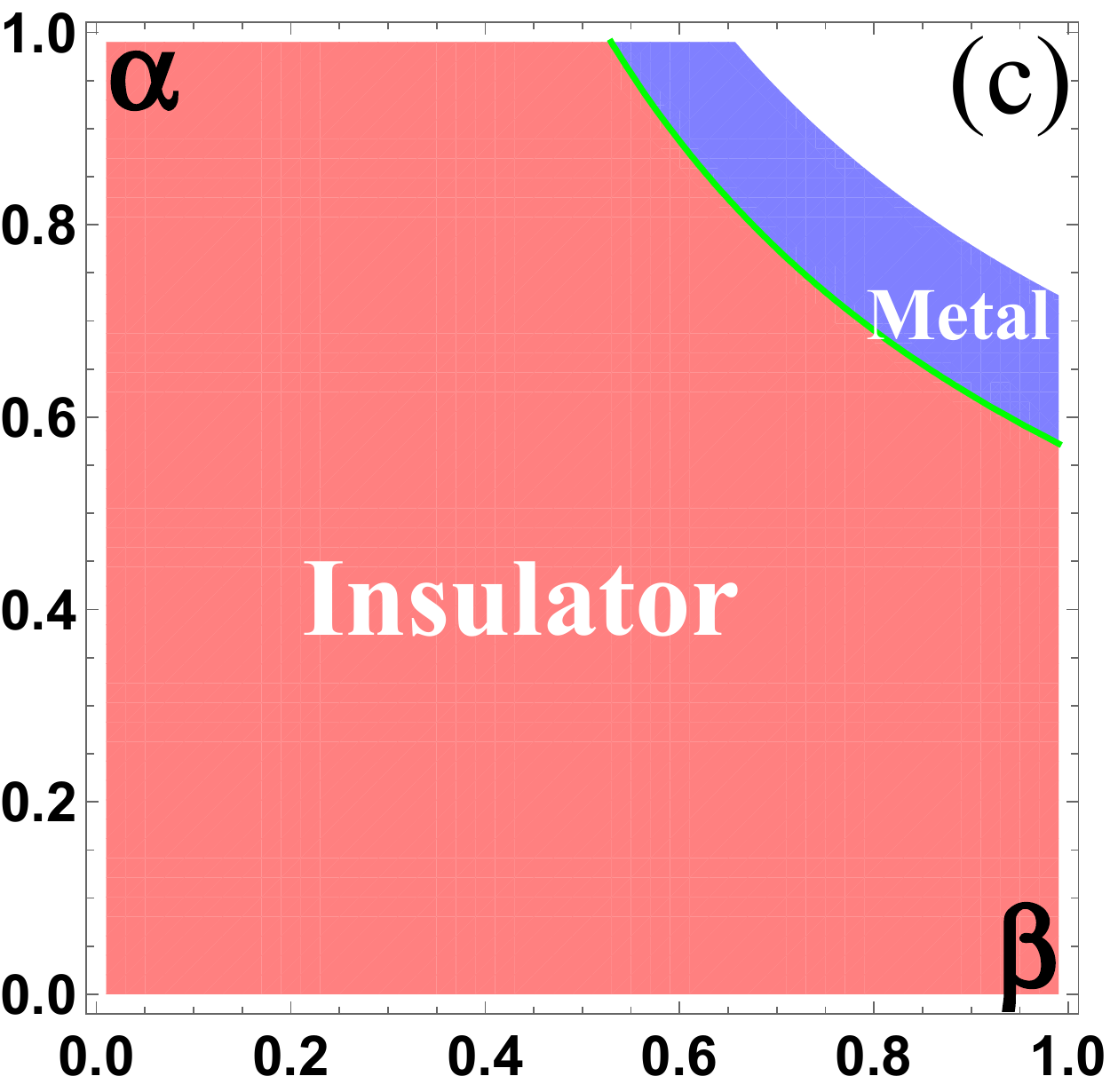}\quad
\includegraphics[width=.15\linewidth]{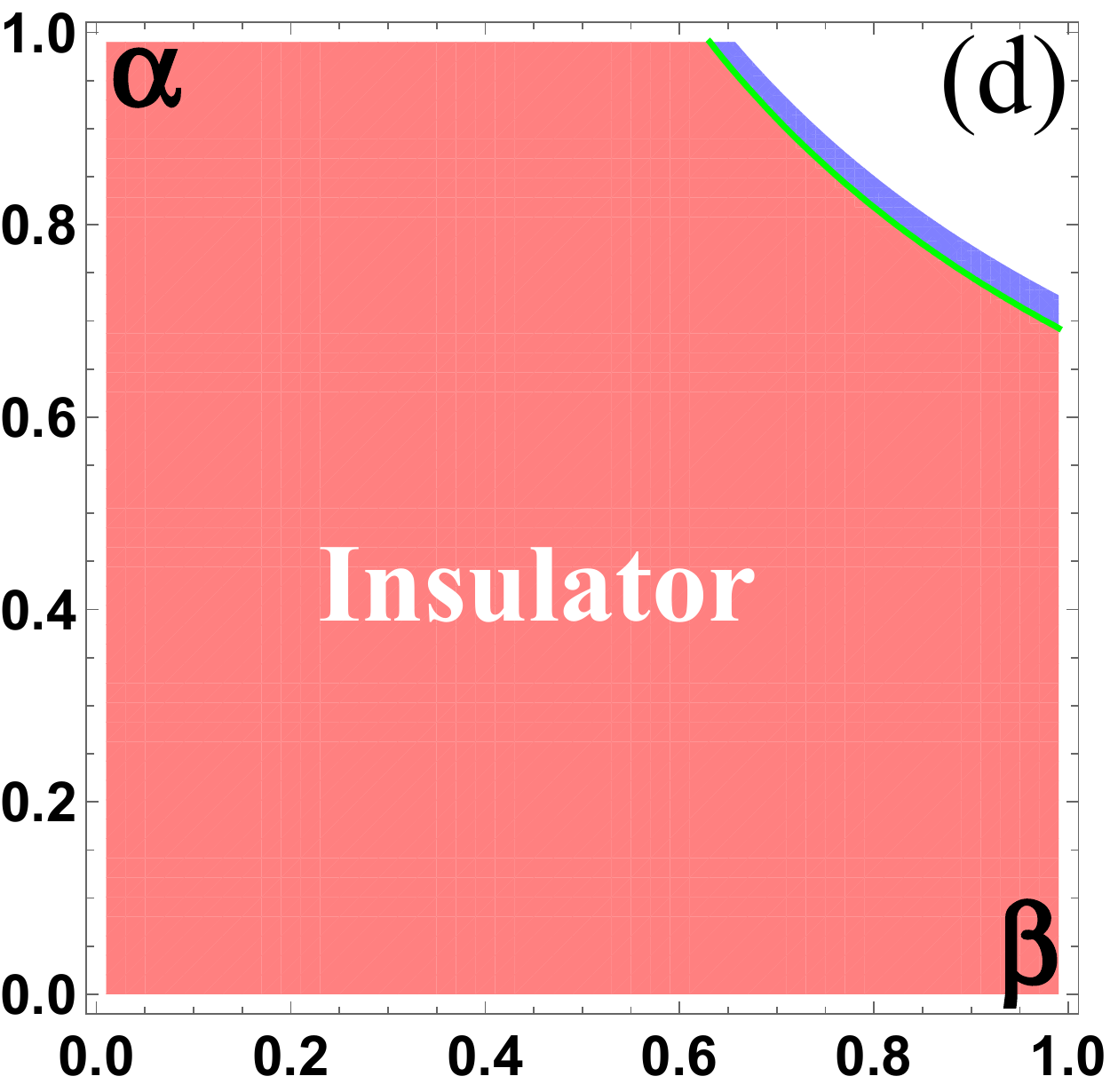}
\caption{(a) ~Scaling dimensions $\Delta_1$ as functions of an  inter-wire interaction strength $\alpha$  and an inter-wire screening parameter $\beta$ for  for three different values of an intra-wire Luttinger parameter $K$ and the transition plane $\Delta_1=1$.
(b-d)~Phase diagrams as cross-sections of Fig. 1a at $\Delta_1=1$ for $K=0.95$, $K=0.75$ and $K=0.55$ correspondingly, exhibiting metal (blue) - insulator (red) transitions.}
\label{fig1}
\end{figure*}


\section{Local impurities}
For a moderate inter-wire interaction the most dangerous perturbation is a single-particle back-scattering within the same wire (for details, see \cite{KLY,IVYnew}). The operator of such perturbation is described by the vectors
${\bf n}_{\mathrm{in}}={\bf n}_{\mathrm{out}}=(0, 0,...,  1_i, 0, ...)^T$, and ${\bf j}_{\mathrm{in}}=-{\bf j}_{\mathrm{out}}=(0, 0,...,  \pm 1_i, 0, ...)^T$ with $i$ labeling the wire, where the scattering takes place. The sign $\pm$ stands for a backscattering of a right- or left-moving particle.  The general expression for the scaling dimension $\Delta_1$ of this process is obtained from Eq.~(\ref{Delta})
\begin{equation}\label{Delta1}
\Delta_1=\frac{K}{K_{\rm ws}}\,,\quad K_{\rm ws}^{-1}=\int_{0}^{\pi}\frac{{\rm d}q}{\pi}\,\frac{1}{r_q}\,.
\end{equation}
The critical value $K_{\rm ws}(\alpha,\beta)$ defines the boundary between metallic and insulating behaviour in the plane interaction parameters $\alpha$ and $\beta$ defined after Eq.~(\ref{scrCoulomb}). At $K>K_{\rm ws}$ the scaling dimension of impurity backscattering is greater than one, $\Delta_1>1$, the impurity is irrelevant at low temperatures, and  corrections to conductance of ideal ballistic system vanish as $\sim(T/D)^{2(\Delta_1-1)}$. At weaker intra-channel repulsion, when $K<K_{\rm ws}$, the scattering is relevant, $\Delta_1<1$, and system behaves as insulator at low temperatures. It is instructive to note here that $K_{\rm ws}<1$ and, therefore, bosons and attractive fermions ($K>1$) are always ideal conductors.

It is important to stress that if a single impurity becomes relevant in some wire, one has to cut this wire and and test the stability of the resulting disjoint configuration against tunneling perturbation connecting two detached semi-infinite wires. In the situation when all but one wires preserve their states (conducting in this case), one can use the duality result derived in \cite{IVYnew} to conclude that the scaling dimension of the tunneling is inversely proportional to $\Delta_1$ and, therefore, only one of the configurations is stable. This is true for a single impurity as well as for a set of impurities separated by a distance greater than the mean free path. In the latter case, all those impurities behave similarly: either all of them will be relevant or irrelevant depending on the interaction strength. Since the impurities are randomly scattered over the whole sample, all wires will be either conducting or insulating. This means that we have a sharp boundary between bulk conducting and insulating states. This situation is very different from what we obtain below for an extended defect.

Our expression Eq.~(\ref{scrCoulomb}) is applicable for any finite-range inter-channel interaction, from the ideal screening ($\beta=0$) to the pure (unscreened) Coulomb one ($\beta=1$). All the phase diagrams contain a metallic state (coloured blue), an  insulating state (red), and the void region, where our description fails ( the Wentzel-Bardeen instability \cite{WB} corresponds to the range of parameters where expression under the square root in Eq.~(\ref{K}) is turning negative).

In Fig.~\ref{fig1}a we show scaling dimensions as functions of the inter-channel parameters $\alpha$ and $\beta$ for three different values of $K$.
Fig.~\ref{fig1}b presents the phase diagram for the case when each wire has a fermionic Luttinger parameter $K=0.95$. When either the inter-wire interaction strength or the screening radius are small ($\alpha\ll 1$ or $\beta\ll 1$ ), any weak back-scattering grows due to the renormalization by interactions, leading to the insulating array, in the complete agreement with Refs. \cite{KF,Scheidl02}. On the other hand, when both strength and radius of the inter-wire interactions increase, the back-scattering in each wire is completely suppressed, and the conducting state is robust. In Figs.~\ref{fig1}(c, d) phase diagrams for $K=0.75$ and $K=0.55$ are presented. It is easy to notice how the boundary between the insulating and conducting states changes with parameter $K$. The smaller $K$ becomes, the more prevailing the insulating state is, once again in the complete agreement with Refs. \cite{KF,Scheidl02}.

\begin{figure*}\vskip 0.5 cm
\centering
\includegraphics[width=0.3 \linewidth]{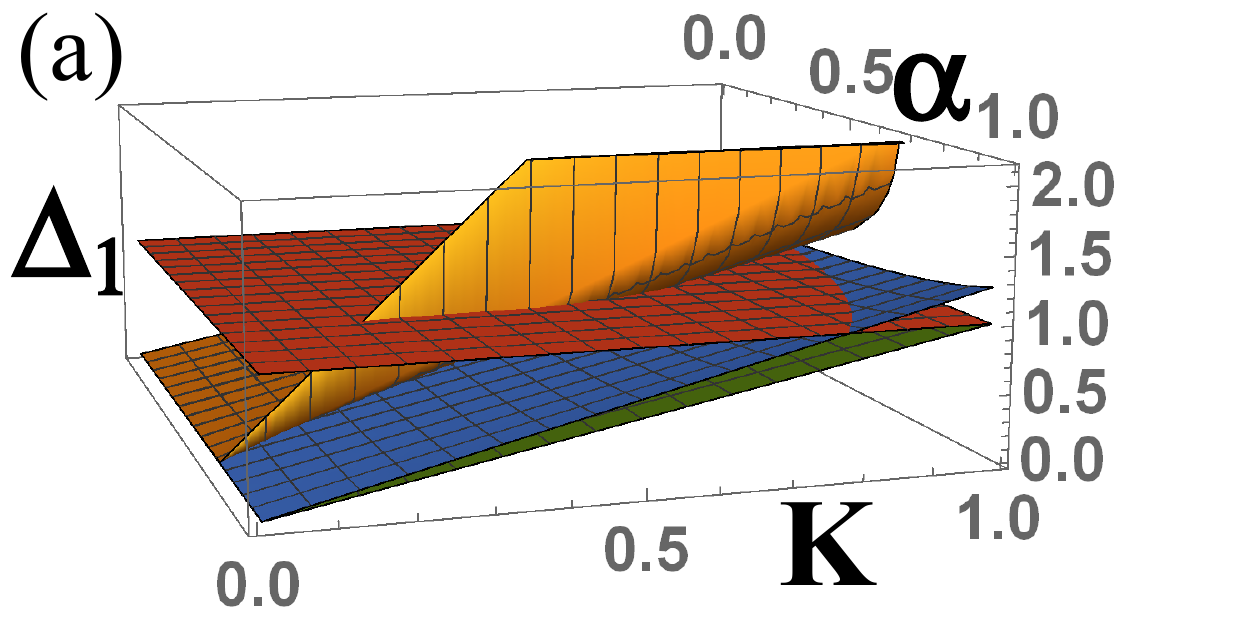}\qquad
\includegraphics[width=0.2 \linewidth]{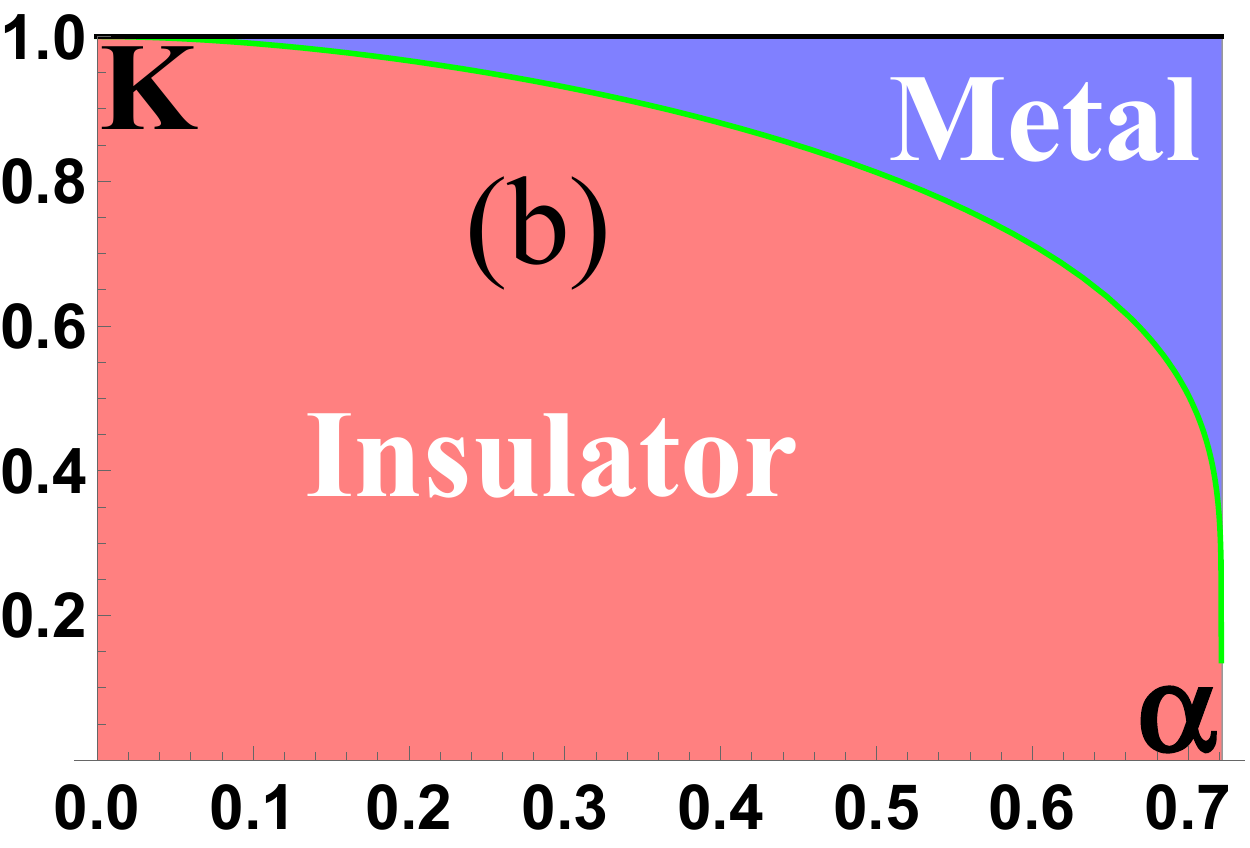}
\includegraphics[width=0.2 \linewidth]{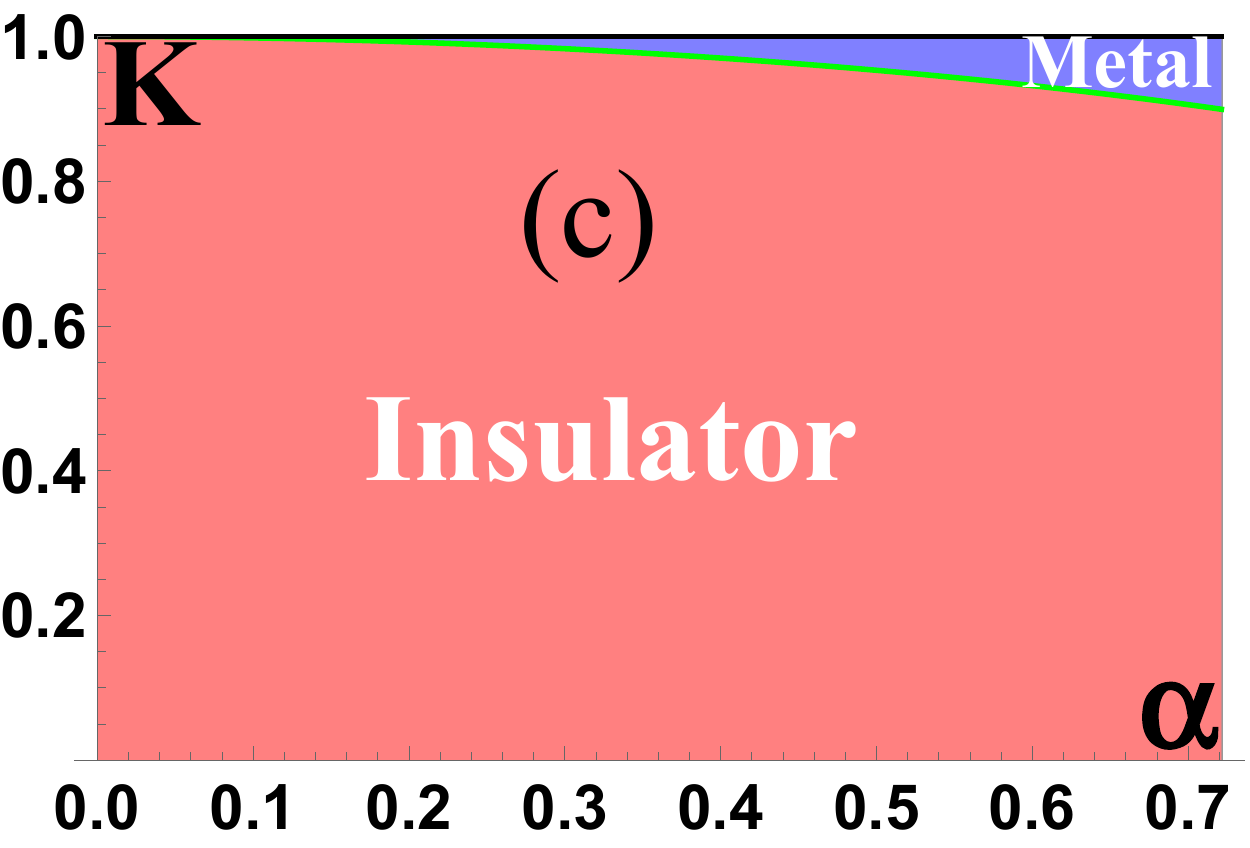}\qquad
\includegraphics[width=0.2 \linewidth]{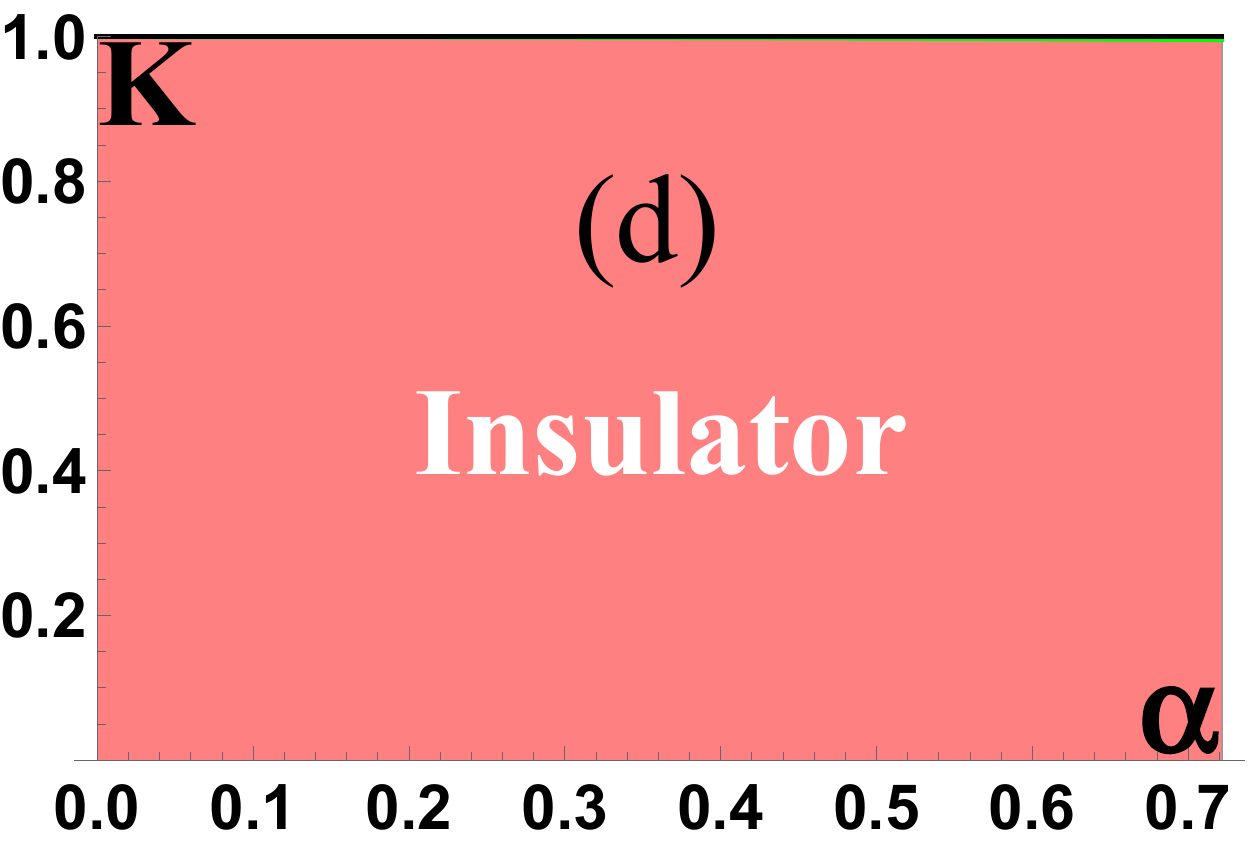}
\caption{(a) ~Scaling dimensions $\Delta_1$ as functions of an  inter-wire interaction strength $\alpha$  and an intra-wire Luttinger parameter $K$ for  for three different values of $\beta$ and  the transition plane $\Delta_1=1$. (b-d)~Phase diagrams  as cross-sections of Fig. 2a at $\Delta_1=1$  for $\beta=1$, $\beta=0.5$ and $\beta=0.1$ correspondingly, exhibiting metal (blue) - insulator (red) transitions.}
\label{fig2}
\end{figure*}

In Fig.~\ref{fig2}a we present scaling dimensions as functions of $\alpha$ and $K$ for three different values of $\beta$. In Fig.~\ref{fig2}b a phase diagram is shown for an infinite inter-wire radius of interactions (parameter $\beta=1$), providing the largest possible phase-space for a conducting state. In  Fig.~ \ref{fig2}c ($\beta=0.5$ ) one can observe the shift of the metal-insulator transition boundary as the transverse screening increases (screening parameter $\beta$ decreases), which reduces the region of the metallic state. Finally, in Fig.~\ref{fig2}d $\beta=0.1$, corresponding to a screening radius of order of half of the inter-wire distance (the inter-wire interaction is almost completely suppressed), the metal state does not exist, as in a single wire.


\section{Line defect}
In what follows we analyze the stability of the sLL phase with respect to an intra-channel back-scattering off a line defect. A line defect is assumed, for simplicity, to be perpendicular to the wires forming sLL (see Fig. \ref{figMixedState}). It can be treated as the multiple backscattering terms present in all wires at the same longitudinal coordinate, $x=0$. All these terms are alike and are  renormalised in a similar fashion. To calculate the scaling dimension of the back-scattering from the weak extended defect, one has to assume that all wires are in the same configuration. In the first approximation with respect to the weakness of the defect, the  scattering operators in different wires are renormalised independently and, therefore, the scaling dimension of each scattering operator is given again by  Eq.{(\ref{Delta1}). The complementary operator of the local weak link (strong defect scattering) has a  scaling dimension found from the inverse of the ${\hat K}$-matrix, as it was derived in \cite{KLY,IVYnew}. Perturbing insulating configuration means that all wires at $x=0$ are cut by the line defect and we insert weak links (tunneling) between the half-wires.  The insulating configuration is, therefore, macroscopically different from the conducting one. To analyse the relevance of weak links we have to use the technique developed in \cite{KLY,IVYnew}. To this end, we need the inverse of the Luttinger matrix which can be found from Eq.~(\ref{K}):
\begin{equation}\label{K-1}
\left[{\hat K}^{-1}\right]_{ij}=\frac{1}{K}\,\int_{0}^{\pi}\frac{{\rm d}q}{\pi}\,r_q\,\cos qn\,,\quad n=i-j\,.
\end{equation}
The scaling dimension of the weak link operators now should be extracted from the Eq.~(\ref{K-1}) since each tunneling is surrounded by the cuts in all other wires. As it was shown in \cite{KLY,IVYnew} the corresponding scaling dimension is equal to the diagonal element of the inverse of the ${\hat K}$-matrix, i.e. Eq.~(\ref{K-1}), with $n=0$. Combining all the results together, we find that two configurations (conducting and insulating) are defined by the two scaling dimensions
\begin{equation}\label{line}
\Delta_{\mathrm{ws}}\equiv\Delta_1= \frac{K}{K_{\rm ws}}\,,\quad
\Delta_{\mathrm{wl}}=\frac{K_{\rm wl}}{K}\,,
\end{equation}
where
\begin{equation}\label{lineK}
K_{\rm ws}^{-1}=\int_0^\pi \frac{{\rm d}q}{\pi}\,\frac{1}{r_{q}}\,,\quad K_{\rm wl}=\int_0^\pi \frac{{\rm d}q}{\pi}\,r_{q}\,.
\end{equation}
As one can see, the line defect analysis is very different from the local impurity one. In the perturbative RG approach, one has to test perturbations around the two possible configurations corresponding to the weak back-scattering and the weak-link. Unlike the local scatterer case, where the two configurations were different by the state of only one wire embedding the scatterer, and the duality relation \cite{IVYnew} guaranteed a single stable fixed point with a sharp boundary between the phases, an extended defect  requires comparison of the configurations where {\em all wires} are either conducting or insulating. The duality is not applicable any longer, which may lead to the existence of two stable and one unstable fixed points and  to a 'split' of the boundary  between the pure states with the appearance of the mixed state region.

Generically, one would expect four different states since either of two scaling dimensions could be above or below unity. Two states where one of the scaling dimensions is above unity and the other one is below are the well known conducting and insulating states. Two states where both scaling dimensions are above or below unity would correspond to the unstable or stable RG fixed points accordingly. In contrast to those expectations,  we found that the inequality $K_{\rm ws}\leq K_{\rm wl}$ is valid for all values of inter-wire interaction parameters, which rules out the existence of a new state described by a stable RG fixed points. As the result, the phase diagram consists of the three states only, as shown in Fig. \ref{figMixedState}) 
\begin{eqnarray}\nonumber
K &<& K_{\rm ws}, \quad \mbox{insulating state}; \\\nonumber
K &>& K_{\rm wl}, \quad \mbox{conducting state}; \\\nonumber
K_{\rm ws} &<& K < K_{\rm wl},   \quad \mbox{mixed state}.
\end{eqnarray}
While the scattering is completely suppressed (enhanced) by interactions in the conducting (insulating) state, the physical picture of the mixed state is more complicated. That state is characterized by both back-scattering and weak tunneling being suppressed by interaction. We conclude, that the behavior of the system depends on the initial (bare) strength of the scattering. At strong bare back-scattering (the weak link condition),  the system ends up in the insulating state at zero temperature. In the contrary, at weak bare back-scattering (the weak scattering condition), the resulting behavior of the system is metallic. This result has a very non-trivial implication for experimental observations. A linear defect in a 2-D array can be created by applying the voltage to  the  perpendicular gate placed underneath the array. If the distance between the gate wire and the array is constant, the linear defect will generate almost equal backscattering in all wires, which will lead to either ideally conducting or insulating states controlled by the gate voltage. If the distance between the gate and the array varies, the scattering strength in some wires may be below a threshold and in others above it. Such an inhomogeneous scattering should lead to a partial suppression of the conductance with few conducting channels. The exact configuration of conducting and insulating wires requires separate study. The conductance observed should become function of the gate voltage distribution.

\section{Summary and conclusion}
In the absence of the inter-wire interaction, any small back-scattering grows due to the renormalization by interactions, leading to the insulating array, in complete agreement with Refs. \cite{KF,Scheidl02}. A finite inter-wire interaction changes this picture though, producing a region of conducting state at small intra-wire interaction (Luttinger parameter $K$ is close to unity), as well as a region of the mixed state at intermediate intra-wire interactions. The existence of the conducting state signals the phenomenon of delocalization by interactions in the array. We note, that the complete analysis of the delocalization requires consistent treatment of multiple impurity scattering, which lies out of the scope of this work. However, the irrelevance of the single-particle back-scattering operator indicates metallic character of transport through the array at low but finite temperatures. In the mixed state, the character of transport can be metallic or insulating, depending on the initial strength of the scattering, since both the weak back-scattering as well as the weak tunneling operators are irrelevant. Therefore, within the state diagram in Fig.~\ref{figMixedState}, one can observe the metal-insulator transition driven by interactions, as well as driven by disorder. Analogous interplay of interactions and disorder has been observed experimentally and confirmed by theoretical RG-calculations \cite{Kravchenko2}.
It should be stressed that the metal-insulator transition driven by a change in either interaction or disorder strength occurs also in 1D disordered LLs. The main difference between the single- and multi-channel situations is that the former transition takes place in the low-temperature region ($T\tau\ll 1$, where the renormalisation of the interaction by disorder should be taken into account, i.e. Kosterlitz-Thouless (KT) phase transition), and in a small vicinity of the Luttinger parameter $K=3/2$ corresponding to attractive fermions. In the latter case, the phase transition can be observed in a finite region of parameters for repulsive fermions in the high-temperature limit, where the renormalisation of interaction can be neglected. The high-temperature states in the mixed state would have been separated by a boundary if we chose to draw a phase diagram with the disorder strength on an extra axis. The RG flows would still be given by straight lines,  unlike the KT RG flows in a single-channel problem.

In conclusion, we analyzed the relevance of the single particle back-scattering in the array of LL wires as a function of intra- and inter-wire interactions. Based on that analysis, we constructed a phase diagram (presented in Figs. \ref{fig1} and \ref{fig2}) for the low-temperature transport through the array. Depending on the strength of interactions, the system exhibits metallic or insulating behavior. Therefore, the transition between the metallic and insulating states in the array can be driven by interactions. We have also analyzed this system in the presence of a linear defect. In this case the array exhibits metallic, insulating and mixed states (presented in Fig. \ref{figMixedState}). The metal-insulator transition in this case can be driven by both interactions and disorder. This behavior is qualitatively similar to the one observed in the experiment on the interaction-driven metal-insulator transition in clean two-dimensional systems \cite{Kravchenko1,Kravchenko2}.

\section*{acknowledgements} IVY research was funded by the Leverhulme Trust Research Project Grant RPG-2016-044.
This research was conducted in frame of the Advanced Study Group at the Center for Theoretical Physics of Complex Systems in the Institute for Basic Science, Daejeon, South Korea. We thank S. Kravchenko for illuminaing discussions.

\end{document}